\documentclass[prb,floatfix,showpacs]{revtex4}
\usepackage{graphicx}
\newcommand{\be}{\begin{equation}}
\newcommand{\ee}{\end{equation}}
\newcommand{\bea}{\begin{eqnarray}}
\newcommand{\eea}{\end{eqnarray}}
\newcommand{\ba}{\begin{array}}
\newcommand{\ea}{\end{array}}
\newcommand{\ben}{\begin{enumerate}}
\newcommand{\een}{\end{enumerate}}
\newcommand{\bei}{\begin{itemize}}
\newcommand{\eei}{\end{itemize}}

\newcommand{\e}{\epsilon}

\newcommand{\om}{\omega}
\newcommand{\Om}{\Omega}

\newcommand{\s}{\sigma}

\newcommand{\cd}{c^\dagger}
\newcommand{\vd}{v^\dagger}

\newcommand{\dek}{\Delta\epsilon_k}

\newcommand{\la}{\langle}
\newcommand{\ra}{\rangle}
\newcommand{\pd}{P^\dagger}

\newcommand{\lm}{\lambda}

\newcommand{\D}{\Delta}

\newcommand{\ua}{\uparrow}
\newcommand{\da}{\downarrow}

\begin{document}
\title{Coherent optical control of correlation waves of spins in semiconductors}
\author{Eran Ginossar, Yehoshua Levinson, and Shimon Levit}
\affiliation{Department of Condensed Matter Physics, The Weizmann Institute of Science, Rehovot 76100, Israel}
\email{eran.ginossar@weizmann.ac.il}
\date{\today}
\begin{abstract}
We  calculate the dynamical fluctuation spectrum of electronic spins in a semiconductor under a steady-state illumination by light containing polarization squeezing correlations.  Taking into account quasi-particle lifetime and spin relaxation for this non-equilibrium situation we consider up to fourth order optical effects which are sensitive to the squeezing phases.
 We demonstrate the possibility to control the spin fluctuations by optically modulating these phases as a function of frequency, leading to a non-Lorentzian spectrum which is very different from the thermal equilibrium fluctuations in n-doped semiconductors. Specifically, in the time-domain spin-spin correlation can exhibit time delays and sign flips originating from the phase modulations and correlations of polarizations, respectively. For higher light intensity  we expect a regime where the squeezing correlations will dominate the spectrum.
\end{abstract}
\maketitle

\section{Introduction}\label{sec-intro}

%

When a semiconductor  absorbs circularly polarized light an
average collective spin polarization $\la \mathbf{S}(t)\ra$ is
induced in the conduction band, a phenomenon known as {\em optical
orientation}
\cite{optical-orientation,Levinson}. In
direct-gap $III-V$ semiconductors this is a result of optical
transitions of electric dipole type across the electron-hole gap.
In this process there is a net transfer of angular momentum from
the helicity of the photons to the  angular momentum of the
electrons. The optical selection rules are such that for the
circularly polarized light the excitation rate for electrons with
one spin projection is larger compared to the other, resulting in
a net spin polarization in the conduction band.  Following the
post excitation spin dynamics one can investigate spin relaxation
mechanisms using techniques such as time-resolved Faraday rotation
and time-resolved photoluminescence
\cite{fabian-rmp,pfaltz}. It is also possible
to monitor the space resolved distribution of spins, their
orientation and magnitude  as well as coherently control these
quantities \cite{awschalom-PD99,awschalom-science01,hagele}.

Next interesting object to study are the fluctuations of the spin
in a semiconductor.  In thermal equilibrium these fluctuations
were measured by employing the Faraday effect \cite{Oestreich-2}.
A linearly polarized probe beam that passes through the sample is
affected by the instantaneous magnetization of the sample, and its
vector of polarization acquires a rotation, proportional to the
magnetization \cite{Landau}. By  measuring the spectrum of the
polarization fluctuations, it is possible to relate it to the spin
fluctuations.

Here we propose to go beyond the equilibrium and to study
dynamical fluctuations of the electronic spins which are induced
by external optical field. In the way similar to the average spin
induced by the polarization  of absorbed light we wish to consider
how fluctuations of the light polarization produce upon absorption
dynamical  fluctuations of electronic spins.  We will consider
polarization squeezed light which has predetermined spectrum of
two photon correlations.   The spin fluctuations spectrum will be
determined by the dynamics of the absorption of such light and
will be sensitive to the {\em phases} of the optical correlation
functions.  These can be controlled opening possibilities for a
coherent control of spin correlations.

In the past theoretical suggestions to observe quantum optical
effects in light-matter interaction involved models with squeezed
radiation reservoirs interacting with atoms and semiconductors
\cite{Gardiner-86,Itay,paper}. This
however  requires a high quality squeezed reservoir, which
experimentally remains challenging to achieve.  Another  quantum
optical effect which does not involve a reservoir is related to
two-photon absorption of squeezed light by a three-level system
\cite{Ficek-1,Ficek-2} and was demonstrated experimentally by
Georgiades \cite{Georgiades}. Other works explored the aspect of
transmission of squeezed light through bulk media
\cite{artoni-loudon,schmidt,beenakker}, and photoionization \cite{kurizki}. Schemes for transferring
correlations from light to matter have recently been developed in
atomic and molecular optics (AMO), both theoretically
\cite{kuzmich,serafini} and experimentally
\cite{hald,lukin}. These schemes employ either coherent
optical dipoles of atomic $V$-systems or ground states coherence
of $\Lambda$-systems, leading to a {\em second order} dependence
of the spin fluctuations on the squeezed optical field.

Unfortunately in semiconductors one is faced with strong dephasing
of optical dipoles as well as valence band spins due to Coulomb,
electron-phonon and spin-orbit interactions, rendering  the above
atomic optics schemes impractical. We note however that the dephasing and relaxation of the conduction band spins are much slower.
In addition continuous energy bands in semiconductors have very different level structure and
optical selection rules. In contrast to the AMO schemes, we
suggest to use this and employ {\em fourth order} optical effects  to
manipulate the collective spin of the conduction band electrons
through the process of two photon absorption.

 For simplicity and to
isolate the transfer of  fluctuations from the average we assume in this work that
the semiconductor is irradiated with the light which is on the
average unpolarized but possesses non vanishing squeezing
correlations between different polarization amplitudes. Such light
(polarization squeezed vacuum) was discussed by Karrasiov
\cite{karrasiov-1}, Lehner \cite{Lehner-96}, Korolkova
\cite{korolkova-2}, and generated in several
experiments \cite{leuchs-1,leuchs-2}. In this paper we demonstrate (see Eq.
(\ref{spin-spin-combined-3})) that by externally manipulating the
phase of the frequency dependent photonic correlations  a new
possibility opens up of optical coherent control of spins in
semiconductors. The essence of this effect lies in controlling the
interference between quantum amplitudes related to many pairs of different
optical frequencies. We predict that in order to observe this effect it is not necessary to
have squeezing in quantum sense, i.e. below the shot noise limit. This should allow for easier measurements
since such light can be generated with higher intensities. Nevertheless we
anticipate that also in the quantum optical regime interesting features should
appear in the spectrum following our previous predictions for the
static spin correlations \cite{paper2}.

The feasibility of observing spin effects in semiconductors relies
on the relatively weak coupling of the conduction band spins to
the environment, i.e. slow spin flipping processes. This has been
experimentally demonstrated in various situations in the past
\cite{fabian-rmp, Awschalom-2, Beschoten, Ohno, pfaltz, Dzhioev}.
We also rely on the fact that for the holes the situation is
different with  the corresponding rates  estimated to be several
orders of magnitude higher
\cite{Levinson,fabian-rmp,optical-orientation}. Accordingly we
have neglected the contribution of the hole spins to the total spin correlations. To observe induced spin
fluctuations it is necessary to extend the experimental
capabilities of measuring spin fluctuations to the non-equilibrium
regime. In principle two beams should be employed in such an
experiment, i.e. a correlated light pump and a linearly polarized
probe acting at the same time. In addition to the measurement of
the induced magnetic moment, it is necessary to have phase
manipulation capabilities in the incoming pump beam, similarly to
those demonstrated recently for squeezed vacuum \cite{dayan}.

In this work we have taken into account scattering effects of
non-radiative processes in a semi-phenomenological way via
relaxation times. However our results do not depend in an essential way on the details of the
interactions and the main qualitative features that we
demonstrate should be reproducible even in a more detailed study. This claim is supported by a separate diagrammatic calculation \cite{mythesis} which reproduces qualitatively the main result that are presented here.

The structure of the paper is as follows. In (\ref{sec-model}) we
introduce the two-band model of electrons interacting with the
driving field. In sections (\ref{sec-1x1},\ref{sec-2x2}) we
calculate the second and fourth order contributions (in the
optical field) to the spin fluctuations, assuming zero lattice
temperature and phenomenological description of relaxations. In
section (\ref{sec-discussion}) we develop physically motivated
simplifications of our results, and discuss their meaning. In
section (\ref{sec-results}) we explore the consequences of phase
modulations of the squeezing correlations on the spin fluctuations
spectrum. In appendix \ref{app-A} we give details concerning the
dipole matrix elements which appear in the light-matter
Hamiltonian. In appendix \ref{app-B} we explain  the physical
nature of our phenomenological approach and how it affects the
results. Finally in appendix \ref{sec-1x3} we discuss the
calculation details of additional fourth order contributions which
are small and not included in the main calculation.


\section{Model}\label{sec-model}
Valence and conduction bands of a direct-gap semiconductor ($III-V$) interacting with the light are modeled by the Hamiltonian
\be\label{H total}
H=\sum_{k\s}\epsilon_{k}^{c}\cd_{k\s}c_{k\s}+\sum_{k'\s}\epsilon_{k'}^v\vd_{k'\s}v_{k'\s}+H_{LM}+H_C+H_{SO}
\ee
where the operators $c_{k\s}$ and $v_{k'\s}$ denote annihilation
operators of the electrons in the conduction and valence bands,
with quasi-momenta $k,k'$ and total angular momentum $\s$. The index $\s$ in the second term enumerates
both the degenerate valence bands and their spin degeneracy. The effect of the
spin-orbit (SO) interaction in the valence bands is usually accounted
for by Luttinger model for degenerate bands \cite{Luttinger-1}. This model takes into account the
mixing of the angular momentum $j=3/2$ states due to spin-orbit
coupling. As we explain in appendix \ref{app-A}, we will take into account the heavy-hole valence band transitions with the energies $\e_k^v$ being  degenerate with respect to the spin projection, and renormalized due to the SO interaction in the valence band.
Generally important are also the Coulomb interaction ($H_C$) between electrons, which is responsible for
excitonic and scattering effects, and spin-orbit interaction in the conduction band ($H_{SO}$). We will focus here on effects related to the light-matter interaction ($H_{LM}$) in the dipole approximation. We study optically excited electrons in
the conduction band with energies above the ionization level of
the exciton. These electrons are well described as quasi-particles
with a finite lifetime arising from the momentum scattering
induced by disorder and Coulomb interactions.  We will
show in appendix \ref{app-A} that to a good approximation it is
possible to write the light-matter interaction as
\be\label{H-LM}
H_{LM}(t)=\sum_{k,p,\s} \left[ d_{k,p}^{\s}\pd_{k,k+p,\s} b_{p,\s}
e^{i\Om_{k,p} t}+d_{k,p}^{\s*}b^*_{p,\s} P_{k,k+p,\s} e^{-i\Om_{k,p}
t} \right] \ee where the operator $P_{k,k+p,\s}=\vd_{k,\s}
c_{k+p,\s}$ is the interband polarization.
$\Om_{k,p}=\e^c_{k+p}-\e^v_k-\om_p$ are the differences between quasi-particle
energies and the photons and $b_{q,\s}$ are optical field amplitudes with wave number $q$ and polarization $\s$. For these transitions it is sufficient to use one index ($\s$), denoting both projection of conduction band spin ($\pm 1/2$),
projection of valence band total angular momentum ($\pm 3/2$), and
helicity ($\pm 1$), depending on the context where it appears. We
also show in appendix \ref{app-A} that for most purposes the
 interaction transition matrix element $d_{k,p}^{s}$ can be
taken as independent of the direction of $k$ and of $s$ and simply
denoted as a constant $d$.

The spin-orbit Hamiltonian ($H_{SO}$) is responsible for the  D'yakonov-Perel'(DP) and Bir-Aronov-Pikus (BAP) spin relaxation mechanisms, which are described by the following effective Hamiltonians\cite{fabian-rmp}

\be\centering  H_{SO}^{(DP)}=\frac{1}{2}\hbar
\sum_{s,s'}(\vec{R}(k) \cdot \vec{\s})_{s,s'}\cd_{ks} c_{ks'} ,
\,\,\,\,\,\,\,\,\, H_{C}^{(BAP)}=A
\sum_{k,k',q}\sum_{s,s',j,j'}(\vec{J} \cdot \vec{\s})_{s,s',j,j'}
\cd_{k,s}c_{k+q,s}\vd_{k',j}v_{k'-q,j'} \ee where $\vec{R}(k)$ and
$A$ are parameters depending on the material, dimensionality, temperature, and details of the optical excitation.
We discuss in section (\ref{sec-1x1}) the main effects of $H_{SO}$ and $H_C$ on the optically excited spin correlations. Note that the Coulomb interaction between electrons in the conduction band ($\cd c\cd c$), which is not included, does not affect the relaxation of the total spin, since it is not interacting with external degrees of freedom.
Even though in principle there are additional spin relaxation mechanisms such as Elliot-Yafet, the dominant mechanisms for spin relaxation are D'yakonov-Perel' and Bir-Pikus\cite{fabian-rmp} when the electron gas is non-degenerate (non-metallic regime). For observing the effects which we discuss here, materials with a relatively long spin lifetime such as GaAs are
advantageous.  Spin relaxation times of the average spin have
been measured by different techniques, for different materials and
experimental conditions\cite{fabian-rmp,Awschalom-2}. These
include n-GaAs quantum wells with different widths\cite{pfaltz},
doping levels\cite{kikkawa}, materials\cite{Beschoten,
Ohno} as well as bulk n-GaAs\cite{Dzhioev}. In these experiments
spin life times ranging from hundreds of picoseconds to tens of
nanoseconds have been reported. In some cases the long life times
were measured even in room temperature\cite{Awschalom-2}. The
contribution of the valence band holes to the driven spin
fluctuations is assumed throughout to be small. This assumption is
realistic in bulk semiconductors from the group III-V, where the
spin flip rates for the holes are up to three orders of magnitude
faster than for the conduction band electrons\cite{fabian-rmp},
with estimates and measurements placing the hole spin lifetime at
around $1ps$. These rates lead to very broad distributions of the
holes spin fluctuations in Fourier space with negligible
contributions near $\om=0$. It is very difficult in experiments to
capture such ($10^3$ GHz) fast oscillations, so in a realistic
experiment we can neglect their effect.

Polarization properties of photons are described by the Stokes parameters \cite{collet-book,korolkova-2}
which in the circular polarization basis are written as time averages of
\be \label{stokes-vec-def} p_i=\sum_{q,q',\lm,\lm'}b_{q\lm}^*(\s_i)_{\lm\lm'} b_{q'\lm'}  \ee
where $\s_{i=0..3}$ denote the unit and Pauli spin matrix. The averaging is over times longer than the typical correlation time of the field. We consider a collinear pump beam with a range of frequencies $\om_0\pm B/2$ above the electron-hole gap and
 time and bandwidth averaged Stokes parameters $\la p_{0}\ra=(2\pi c/LB)\sum_{q\lm}N_{q\lm}$,
$\la p_{1}\ra =\la p_{2}\ra=\la p_{3}\ra =0$, where $N_{q\lambda}$ is the average photon occupation per mode and
$L$ is the mode quantization length. This is unpolarized light with the Stokes vector fluctuating around the origin of the Poincar\'{e} sphere.
The fluctuations are described by the covariance matrix $p_{ij}=\la p_i p_j \ra$, which for a Gaussian type field depends
on the normal $\la b^*_{q\lm} b_{q'\lm'}\ra$ as well as anomalous $\la b_{q\lm} b_{q'\lm'}\ra$ correlations, the latter constituting the main characteristics of squeezed light \cite{collet-loudon}. 
In addition to normal correlations  $\la b^*_{q\lm}b_{q'\lm'} \ra=N_q\delta_{\lm \lm'}\delta_{q q'}$ they possess four generic anomalous correlations:
 two for the same polarization squeezing $\la b_{q\pm}b_{q'\pm} \ra=M^{(1)}_{q\pm}\delta_{q+q',2q_0}\delta_{\om_q+\om_{q'},2\om_0}$
and two for the opposite polarization squeezing $\la b_{q\pm}b_{q'\mp} \ra=M^{(2)}_{q\pm}\delta_{q+q',2q_0}\delta_{\om_q+\om_{q'},2\om_0}$,
where $M^{(1,2)}_{q\pm}$ are complex functions and $\om_0=c q_0$. It can be shown \cite{dalton-review,ficek-drummond} that for quantum fields $|M^{(1,2)}_{q\lm}|\le N_{q\lm}\sqrt{N_{q\lm}+1}$ while $|M^{(1,2)}_{q\lm}|\le N_{q\lm}$ for classical fields, which we discuss here. We will later remark on the possible effects of the quantum regime in section \ref{sec-results}. This type of light is also called polarization-squeezed-vacuum \cite{korolkova-2}, or two-mode squeezed vacuum state of type II\cite{kwiat,Lehner-96}.

\begin{figure}\centering
\includegraphics[scale=0.7]{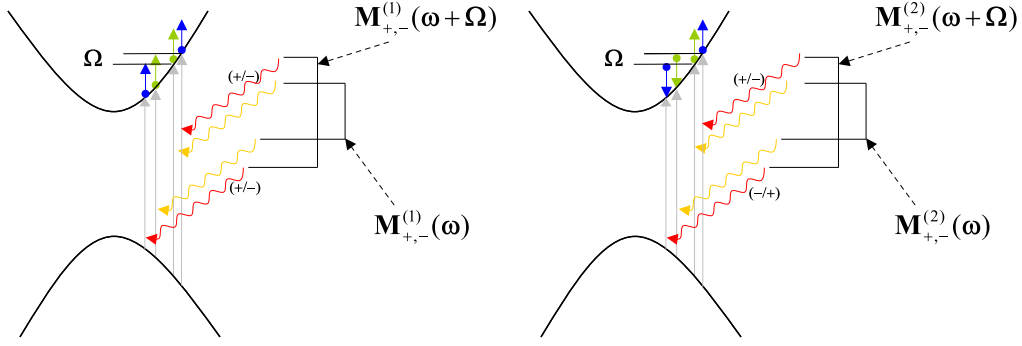}
\caption{(Color online) Illustration of double-photon excitation processes (long grey arrows), with
yellow and red colors for the two pairs, which excites spin-polarized electrons (short arrows). The left (right) figure describes absorption amplitudes of photons of same (opposite) helicity. The interference of these amplitudes creates a spin correlation wave with a frequency $\Omega$.}
\label{dynamical-flucs-figure}\end{figure}

We begin by deriving the Heisenberg equations of motion for the spin
waves in the conduction band. The spin density operator\cite{fetter-book} is given by
\be \vec{S}(r,t)=\sum_{s,s'}
\psi^\dagger_s(r,t) \vec{\s}_{s,s'} \psi_{s'}(r,t)
=\sum_{s,s'}\sum_{k,k'}e^{-i(k-k')r}\cd_{k,s}(t)\vec{\s}_{s,s'}
c_{k',s'}(t)\ee and we define the $q$-component of the spin as \be
\vec{S}(q,t)=\int_{V}d^3r e^{-iqr} \vec{S}(r,t)=\sum_{k,s,s'}
\cd_{k-q,s}(t)\vec{\s}_{s,s'} c_{k,s'}(t) \ee

Without the optical excitation, the spin correlations are zero when the lattice temperature is  $T=0$,
because there are no electrons in the conduction band. When the electric field of the light is stochastic with a zero average, so is the average spin component $\la \vec{S}(q,t)\ra$, and therefore we study the correlations. As we will
discuss in sections \ref{sec-2x2}, due to the optical selection
rules only the averages $\la S_z(q,t)S_z(q',t')\ra$, with  $S_z(q,t)=\sum_k \left[\cd_{k-q,\uparrow}c_{k,\uparrow}-\cd_{k-q,\downarrow}c_{k,\downarrow}\right]$, are affected
by the squeezing correlations of the light beam (directed along the $\hat{z}$ direction) and therefore we will
focus on them. The function $\la S_z(q,t)S_z(q',t')\ra$ is related to experiment in a similar way
that the dipole fluctuations $\la \s(t)_+\s(t')_-\ra$ of an atom are related to the measurable fluorescence spectrum\cite{Walls-book} through the first order correlation function of the optical fields\cite{Glauber-1}, assuming we have a suitable spectrometer at hand. In our case of the spins one way that comes to mind is making use of the Faraday effect, which is customarily used to measure optically injected spin in semiconductors.

Using the Hamiltonian (\ref{H total},\ref{H-LM}) the equations of motion which define the spin waves
are given by \bea \label{eoms-basic} && \frac{d}{dt}
n_{k,k',\s}^c=i\delta\e^c_{k,k'}\cd_{k,\s} c_{k',\s}+i\sum_p
\left[-d \pd_{k'-p,k,\s}b_{p\s}+ d^*b^*_{p\s}P_{k-p,k',\s}\right]
+i[H_{SO},n_{k,k',\s}^c]+i[H_C,n_{k,k',\s}^c] \\ \nonumber &&
\frac{d}{dt} n_{k,k',\s}^v=i\delta\e^v_{k,k'}\vd_{k,\s}
v_{k',\s}+i\sum_p \left[ d \pd_{k',k+p,\s}b_{p\s}-
d^*b^*_{p\s}P_{k,k'+p}\right]
+i[H_{SO},n_{k,k',\s}^v]+i[H_C,n_{k,k',\s}^v]\\ \nonumber &&
\frac{d}{dt}\pd_{k',k,\s}=\frac{d}{dt}\cd_{k,\s}
v_{k',\s}=i\Delta\e_{k,k'}\cd_{k,\s} v_{k',\s}-id^*\sum_p
b^*_{p\s}\left[
n^c_{k,k'+p,\s}-n^v_{k-p,k',\s}\right]+i[H_{SO},\pd_{k',k,\s}]+i[H_C,\pd_{k',k,\s}]
\\ \nonumber
\eea with $\delta\e^c_{k,k'}=\e^c_k-\e^c_{k'}$,$\Delta
\e_{k,k'}=\e^c_k-\e^v_{k'}$, and $n^{c}_{k,k',\s}=\cd_{k,\s}
c_{k',\s}$  \footnote{Note the difference in notation between
$P_{k,k'}$ and $n^{(c,v)}_{k,k'}$. For $P_{k,k'}$ the first index
is always the index of $v_k$ and for $n^{(c,v)}_{k,k'}$ the first
index is the index of $\cd$ or $\vd$.}.

We treat the interaction between the electrons and the optical field in perturbation
theory. The spin operator in Heisenberg picture can formally be written as
\be S_z(q,t)=S_z(q,t)^{(0)}+S_z(q,t)^{(1)}+S_z(q,t)^{(2)}+S_z(q,t)^{(3)}+S_z(q,t)^{(4)}+...\ee

We shall see in section \ref{sec-1x1} that to second order the spin fluctuations are proportional only to the photon occupation $<b^*b>$, and therefore are not affected by squeezing, where latter only makes an effect in the fourth order.
When using it to expand $\la S_z(q,t)S_z(q',t') \ra$ we have
to take all the possible combinations which contain the
optical interaction an even number of times, since the spin operators
conserve the number of particles \bea \nonumber && \la
S_z(q,t)S_z(q',t')\ra=\la S_z^{(0)}(q,t)S_z^{(0)}(q',t')\ra+\la
S_z^{(1)}(q,t)S_z^{(1)}(q',t')\ra+ \\ \nonumber && +\la
S_z^{(0)}(q,t)S_z^{(4)}(q',t')\ra+\la
S_z^{(4)}(q,t)S_z^{(0)}(q',t')\ra+\la
S_z^{(1)}(q,t)S_z^{(3)}(q',t')\ra+ \\ \nonumber
&& +\la S_z^{(3)}(q,t)S_z^{(1)}(q',t')\ra +\la S_z^{(2)}(q,t)S_z^{(2)}(q',t')\ra\\
\eea
However, at $T=0$ the combinations $0\times 0,~4 \times 0,~0\times 4$ are zero since the
spin operator (on the "0" side) would be acting on the empty conduction band state. Therefore we are left with
\bea \label{correlation-genenral-expansion}
&& \la S_z(q,t)S_z(q',t')\ra=\la S_z^{(1)}(q,t)S_z^{(1)}(q',t')\ra+\la S_z^{(2)}(q,t)S_z^{(2)}(q',t')\ra+ \\ \nonumber
&& +\la S_z^{(1)}(q,t)S_z^{(3)}(q',t')\ra+\la S_z^{(3)}(q,t)S_z^{(1)}(q',t')\ra
\eea
i.e. only 2nd and 4th order contributions. The former are not affected by squeezing and contribute to the background of the fluctuations spectrum.
The latter, however, is affected by the q-dependent phase modulations of the function $M^{(1,2)}_{q\pm}$, through a microscopic process of double absorption (see section \ref{sec-2x2}). This process is illustrated in Fig. \ref{dynamical-flucs-figure} showing two cases of opposite and same polarization. Two quantum amplitudes related to absorption of two different pairs of photons proportional to $M^{(1,2)}(\om)$ and $M^{(1,2)}(\om+\Om)$ will be shown to appear in interference terms like $M^*(\om)M(\om+\Om)$ which bring into effect their externally controlled phase difference.

We shall derive the Heisenberg equations of motion up to 4th order in the optical field, by writing formal solutions and substituting back to the previous order. Then we will use the resulting expansion to calculate the averages in (\ref{correlation-genenral-expansion}) order by order, taking into account the nature of the optical field. An important part of the model is the treatment of electron and spin lifetimes, which we will discuss in section (\ref{sec-1x1}).

\section{Calculation of the second order contribution}\label{sec-1x1}

This physical optical process, of the lowest order, is responsible for generating carrier and spin densities in the conduction band of a semiconductor. It is convenient to analyze their space-time profile in Fourier space ($q,\om$), given by the correlation function of $S_z(q,\om)$ operators. The perturbative correction which we discuss here turns out not to depend on squeezing or coherent properties of the light field, and will ultimately just serve as a background for the more interesting higher order processes ($2 \times 2$). At the last stage of the calculation (see Eq. (\ref{spin-spin-2nd-order-3})) we neglect the momenta of the spin wave ($q$) and the photon ($p$) compared with the typical electronic momentum ($k$), in places where they appear together as a sum (e.g. $k+p \rightarrow k$). Within this approximation the only angular dependence that remains is that of the the dipole matrix element (see appendix \ref{app-A}), which amounts to renormalization of  the value of the matrix element.
Now we use (\ref{eoms-basic}) to construct the equation for the spin, with $b_{q,\ua}$ ($b_{q,\da}$) denoting right (left) polarizations, respectively. The first equation then reads
\bea
\label{s_z eom}
&& \frac{d}{dt}S_z(q,t)=\sum_k i\delta\e_{k-q,k} e^{i\delta\e_{k-q,k} t}(n^c_{k-q,k,\ua}-n^c_{k-q,k,\da})+\sum_k\left[i\sum_p\left( -d \,\pd_{k-p,k-q,\ua}b_{q\ua}+d^*b^*_{p\ua}P_{k-p-q,k,\ua}\right)- \right.\\ \nonumber
&&\left.  -i\sum_p\left( -d \,\pd_{k-p,k-q,\da}b_{p\da}+d^*b^*_{p\da}P_{k-p-q,k,\da}\right)\right] +i[H_{SO},S_z(q,t)]+i[H_{C},S_z(q,t)]
\eea
The first term on the r.h.s describes a free evolution of the spin wave with typical frequencies of $v_0 q$, where $v_0$ is the electron velocity at $k_0$, the latter being the quasi-momentum of the electron optically excited by the photon of the middle frequency $\om_0$. Since the light induces spin correlations with the dispersion of $c q$, we are physically motivated to neglect the slow free evolution ($v_0 q$) in the following treatment.

Instead of developing a microscopic theory for the relaxation of the spin wave $S_z$ and interband polarization $P$, we replace the combined effect of $H_{SO}$ and $H_C$ by a quasi-particle description with a width $\gamma=\tau^{-1}$ and a phenomenological relaxation $\gamma_s=\tau_s^{-1}$. Under broadband excitation conditions, and when the electrons kinetic energy is large compared to the electron-hole exchange energy, it is reasonable to model them as quasi-particles with a finite life time ($\tau$). This lifetime enters the model through the {\em level width} of the electronic energies $\e^c_k$. In addition to $\tau$ it is physically plausible, and supported my many experiments, to assume the existence of another, {\em macroscopic} relaxation time ($\tau_s$) of the spin wave which is usually much longer than the electron lifetime, $\tau_s \gg \tau$ for semiconductors such as GaAs. The spin lifetime can be entered as a phenomenological decay term ($-\gamma_s S_z$) in the equation of motion for the spin operator $S_z$ (Eq. (\ref{s_z eom})), similarly to a decay term in a quantum-Langevin equation of motion. Equivalently, it can enter as an additional imaginary part (broadening) to the frequency $\om_q$ of the spin wave. We explain in appendix \ref{app-B} the limitations of the phenomenological approach, and how to reconcile it with the results that we obtained for the static correlations in the previous work \cite{paper2}. The averaging of $\la S_z(q,t)S_z(q',t')\ra$ over the ground state of with $T=0$ leaves only the terms $\la P \pd \ra$. The general expression reads
\bea \label{spin-spin-2nd-order-1}
&& \la S_z(q,t)S_z(q',t')\ra^{(2)}= \\ \nonumber
&& =|d|^2e^{-\gamma_s (t+t')}\int_{t_0}^t dt_1 e^{\gamma_s t_1} \int_{t_0}^{t'} dt_2
e^{\gamma_s t_2} \sum_{k,k'}\sum_{p,p',\s,\s'} (-1)^{\s-\s'} e^{i(\om_p t_1-\om_{p'}t_2)}\la b_{p\s}^* b_{p'\s'}\ra \la P_{k-p-q,k,\s}(t_1) \pd_{k'-p',k'-q',\s'}(t_2)\ra \eea
where $t_0$ is the initial time when the system was in the ground state before the optical fields were turned on.
Given that $N_{p\s}$ is the photon occupation function, $\gamma$ is the quasi-particle lifetime, and $\s,\s'=\pm 1/2$, we also have
\bea \label{model-correlations-for-ss2}
&& \la b_{p\s}^* b_{p'\s'}\ra = \delta_{p,p'}\delta_{\s,\s'} N_{p\s} \\ \nonumber
&& \la P_{k-p-q,k,\s}(t_1)\pd_{k'-p',k'-q',\s'}(t_2)\ra=\delta_{\s,\s'}\delta_{k-p-q,k'-p'}\delta_{k,k'-q'}e^{-i\Delta\e_{k,k-p-q}(t_1-t_2)}e^{-\gamma|t_1-t_2|}.
\eea

Working with this expression it is straightforward to get for $t'>t$, and steady state $t-t_0 \gg \gamma^{-1},\gamma_s^{-1}$
\bea \label{spin-spin-2nd-order-2}\nonumber
&& \la S_z(q,t)S_z(q',t')\ra^{(2)}=|d|^2\delta_{q,-q'}\sum_{k,p} N_p \left[\frac{e^{-(i\Om_{k,p,q}+\gamma)(t'-t)}}{\gamma_s^2-(\gamma+i\Om_{k,p,q})^2}-\frac{\gamma}{\gamma_s}
\frac{e^{-\gamma_s(t'-t)}}{(\gamma_s-i\Om_{k,p,q})^2-\gamma^2}\right] \\ \eea
where $N_p=\sum_{\s}N_{p,\s}$ and $\Om_{k,p,q}=\om_p-\e^c_k+\e^v_{k-p-q}$. Note that since $d$ is the interaction matrix element of the light-matter interaction, it scales like $V^{-1/2}$ because of the electric field normalization. Therefore the correlations scale with the volume as $V$ and are dimensionless (with our definition of $S_z$).
It is instructive to study the structure of the expression inside the brackets in Eq. (\ref{spin-spin-2nd-order-2}). First we note that it is stationary since the electronic correlations in Eq. (\ref{model-correlations-for-ss2}) and the optical fields are also stationary.
For a fixed $k$ and $p$ we expect naturally for the frequency $\Om_{k,p,q}$ to appear with $t-t'$ (the first term). However we also expect a buildup of a constant carrier density in the conduction band, which carries an inevitable static spin density fluctuation $\la S_z^2\ra$. Therefore we expect the spin fluctuation to have a DC Fourier component (the second term), balanced by the spin relaxation $\gamma_s$, and this indeed will turn out to be the case when we evaluate the $k$-summation (see below section \ref{sec-discussion}).

\section{Effects of squeezing}\label{sec-2x2}

\begin{figure}\centering
\includegraphics[scale=0.7]{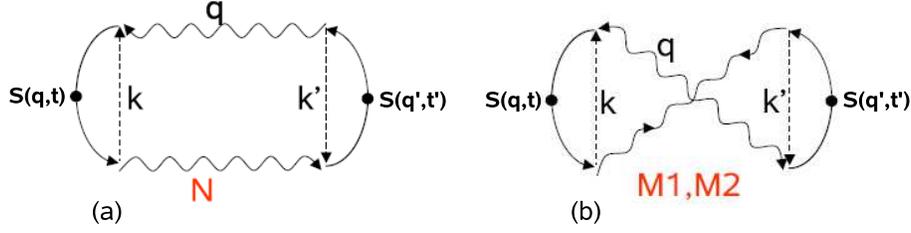}
\caption{A diagrammatic representation of the resonant interaction processes in the normal and anomalous channels leading to spin correlations. The normal channel (a) is related to the auto-correlation function $N$ of optical modes with themselves, whereas the anomalous channel (b) involve a correlation between different optical modes, either of same polarization ($M_1$) or of opposite polarization ($M_2$).}
\label{diagrams-sketch}\end{figure}

Let us now turn to the fourth order contribution to the correlations. Using expression (\ref{sz-second-order-explicit}), we first average over the electronic $T=0$ ground state of the semiconductor (denoted by $\la \cdot \ra_{el}$). This averaging can be done on $S_z(q,t)^{(2)}$ independent of $S_z(q',t')^{(2)}$ since for $T=0$ there are no connected parts between the operators $n^{c,v}_{k,k'}$ in $\la S_z(q,t)^{(2)}S_z(q',t')^{(2)}\ra$. This gives, for steady state ($t-t_0 \gg \gamma^{-1},\gamma_s^{-1}$)
\be \la S_z(q,t)^{(2)}\ra_{el}= |d|^2\sum_{k,p}\left[\mathcal{S}_{k,p,\s=\ua}-\mathcal{S}_{k,p,\s=\da}\right]\ee
with
\be \mathcal{S}_{k,p,\s}=\frac{e^{-i(\om_p-\om_{p-q})t}b^*_{p-q\s}b_{p\s}}{(\gamma_s-i(\om_p-\om_{p-q}))(\gamma-i(\D\e_{k-q,k-p}-\om_{p-q}))}+  \frac{e^{i(\om_p-\om_{p+q})t}b^*_{p\s}b_{p+q\s}}{(\gamma_s+i(\om_p-\om_{p+q}))(\gamma+i(\D\e_{k,k-p-q}-\om_{p+q}))}\ee
where we introduce $\gamma,\gamma_s$ already at the operator level as in the Heisenberg-Langevin approach, this being mathematically equivalent to adding them at the end as representing a single particle level broadening and spin wave frequency broadening. For the spin correlations we have the following dependence on fourth order correlations of the photon fields ($\la \cdot\ra$ denotes full averaging)
\bea \label{spin-spin-4th-1}
&& \la S_z(q,t)S_z(q',t') \ra^{(4)}=|d|^4\sum_{k,p}\sum_{k',p',\lm}\left[\right. \\ \nonumber
&& \frac{e^{-i(\om_p-\om_{p-q})t}e^{-i(\om_{p'}-\om_{p'-q'})t'}
[\la b^*_{p-q\lm}b_{p\lm}b^*_{p'-q'\lm}b_{p'\lm}\ra-\la b^*_{p-q\lm}b_{p\lm}b^*_{p'-q'\bar{\lm}}b_{p'\bar{\lm}}\ra]}
{(\gamma_s-i(\om_p-\om_{p-q}))(\gamma-i(\D\e_{k-q,k-p}-\om_{p-q}))
(\gamma_s-i(\om_{p'}-\om_{p'-q'}))(\gamma-i(\D\e_{k'-q',k'-p'}-\om_{p'-q'}))}+ \\ \nonumber
&& \frac{e^{i(\om_p-\om_{p+q})t}e^{i(\om_{p'}-\om_{p'+q'})t'}
[\la b^*_{p\lm}b_{p+q\lm}b^*_{p'\lm}b_{p'+q'\lm}\ra-\la b^*_{p\lm}b_{p+q\lm}b^*_{p'\bar{\lm}}b_{p'+q'\bar{\lm}}\ra]}
{(\gamma_s+i(\om_p-\om_{p+q}))(\gamma+i(\D\e_{k,k-p-q}-\om_{p+q}))
(\gamma_s+i(\om_{p'}-\om_{p'+q'}))(\gamma+i(\D\e_{k',k'-p'-q'}-\om_{p'+q'}))}+ \\ \nonumber
&& \frac{e^{-i(\om_p-\om_{p-q})t}e^{i(\om_{p'}-\om_{p'+q'})t'}
[\la b^*_{p-q\lm}b_{p\lm}b^*_{p'\lm}b_{p'+q'\lm}\ra-\la b^*_{p-q\lm}b_{p\lm}b^*_{p'\bar{\lm}}b_{p'+q\bar{\lm}}\ra]}
{(\gamma_s-i(\om_p-\om_{p-q}))(\gamma-i(\D\e_{k-q,k-p}-\om_{p-q}))
(\gamma_s+i(\om_{p'}-\om_{p'+q'}))(\gamma+i(\D\e_{k',k'-p'-q'}-\om_{p'+q'}))}+ \\ \nonumber
&& \left. \frac{e^{i(\om_p-\om_{p+q})t}e^{-i(\om_{p'}-\om_{p'-q'})t'}
[\la b^*_{p\lm}b_{p+q\lm}b^*_{p'-q'\lm}b_{p'\lm}\ra-\la b^*_{p\lm}b_{p+q\lm}b^*_{p'-q'\bar{\lm}}b_{p'\bar{\lm}}\ra]}
{(\gamma_s+i(\om_p-\om_{p+q}))(\gamma+i(\D\e_{k,k-p-q}-\om_{p+q}))
(\gamma_s-i(\om_{p'}-\om_{p'-q'}))(\gamma-i(\D\e_{k'-q',k'-p'}-\om_{p'-q'}))} \right] \\ \nonumber
\eea
where each photon correlator can be factorized into products using intensity $\la b^*_{q_1,\lm_1} b_{q_1,\lm_1}\ra$ and squeezing $\la b_{q_1,\lm_1} b_{q_2,\lm_2}\ra$ since the field distribution is Gaussian for light which is down-converted using a $\chi^{(2)}$ non-linearity \cite{collet-loudon}.
As in the case of static spin correlations\cite{paper2}, we assume a symmetric distribution of the correlations as functions of wave vectors, i.e. $q+q'=2p_0$, which holds for non-degenerate (broad spectrum) down-converted light\cite{dalton-review}. We can write for the correlations
\bea  \label{four-fields-correlations}
&& \la b^*_{p-q,\ua}b_{p,\ua}b^*_{p',\ua}b_{p'-q',\ua} \ra =N_{p-q} N_p\delta_{q,q'}\delta_{p,p'}+M^{(1)*}_{p-q}M^{(1)}_p\delta_{q,q'}\delta_{p'+p-q,2p_0} \\ \nonumber
&& \la b^*_{p-q,\ua}b_{p,\ua}b^*_{p'+q',\ua}b_{p',\ua} \ra =N_{p-q} N_p\delta_{q,q'}\delta_{p-q,p'}+M^{(1)*}_{p-q}M^{(1)}_p\delta_{q,q'}\delta_{p'+p,2p_0} \\ \nonumber
&& \la b^*_{p,\ua}b_{p+q,\ua}b^*_{p',\ua}b_{p'-q',\ua} \ra =N_p N_{p+q}\delta_{q,q'}\delta_{p+q,p'}+M^{(1)*}_{p}M^{(1)}_{p+q}\delta_{q,q'}\delta_{p'+p,2p_0} \\ \nonumber
&& \la b^*_{p,\ua}b_{p+q,\ua}b^*_{p'+q',\ua}b_{p',\ua} \ra =N_p N_{p+q} \delta_{q,q'}\delta_{p',p}+M^{(1)*}_{p}M^{(1)}_{p+q}\delta_{q,q'}\delta_{p'+p+q,2p_0}
\eea
where $p_0$ is the central wave-vector of the optical spectrum, and note we omitted the $\pm$ index assuming $M^{(\alpha)}_{q+}=M^{(\alpha)}_{q-}$ for simplicity. In figure (\ref{diagrams-sketch}) we can see a diagrammatic representation of the microscopic processes that lead to the `N' and `M' terms in the expression (\ref{four-fields-correlations}). These are essentially two particle-hole excitations which are correlated through the existence of photonic correlations. These photon correlations are spectral functions, which
can be externally controlled at the source \cite{korolkova-2,leuchs-1,leuchs-2}.
Given these functions it is possible to characterize the fluctuations of the Stokes vector, whose average is assumed to be zero in this calculation (unpolarized) \cite{paper2}.


We note that there is also a $\la S_z(q,t)^{(1)}S_z(q',t')^{(3)}\ra$ contribution which we discuss in detail in appendix \ref{sec-1x3}. This contribution has a much smaller phase space compared to $\la S_z(q,t)^{(2)}S_z(q',t')^{(2)}\ra$ and therefore we neglect it. Also, we have shown\cite{mythesis} that a direct diagrammatic evaluation of the 4th order diagrams, Fig. \ref{diagrams-sketch}, done in the Keldysh two-time formalism gives a result similar to expression (\ref{spin-spin-4th-1}), with the only difference being the replacement $\gamma_s \rightarrow \gamma$. The reason for this is that in the microscopic calculation we have not taken into account explicitly the spin relaxation mechanisms, which is a topic for further research.

\section{Discussion and Results}\label{sec-discussion}

\begin{figure}\centering
\includegraphics[scale=0.5]{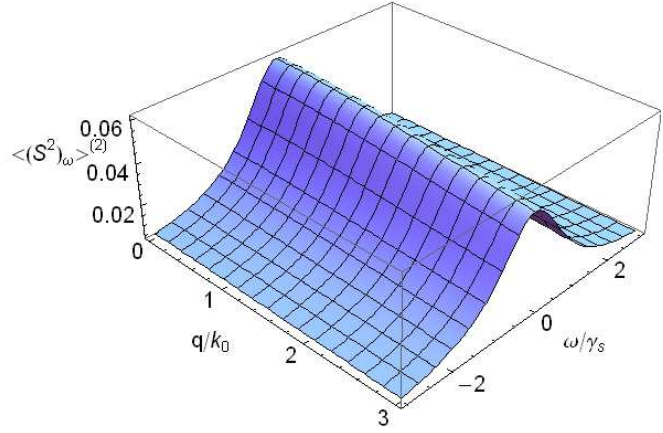}
\caption{The spin fluctuations spectrum (second order contribution) plotted as a function of $(q,\om)$ for excitations with a wave vector parallel to a unidirectional beam, scaled with the central electronic wave vector $k_0$ and the spin relaxation rate $\gamma_s$, respectively. }
\label{E2 spectrum fig}\end{figure}

We would like now to discuss and simplify the expressions for the spin-spin correlations obtained above.
We begin from the second order contribution (\ref{spin-spin-2nd-order-2}) which is quite a complicated integral when the vector nature of $k,p,q$ is taken into account. First note that for the problem we have in mind $p \ll k$, since typically $p\sim 1 \mu m^{-1}$ whereas $k_0\sim 30 \mu m^{-1}$ for electrons with kinetic energy of $30meV$. In Fig. (\ref{E2 spectrum fig}) results of numerical evaluation are shown for the time Fourier transform of Eq. (\ref{spin-spin-2nd-order-2}), assuming a unidirectional beam $\vec{p}||\hat{z}$ and spin wave vectors in the direction of the beam $\vec{q}||\vec{p}$.
We see that the $q$-dependence is very slow for $q\ll k_0$, and similarly it can be shown that the integrand dependence on $p$ is negligible for $p \ll k_0$. Therefore a reasonable zeroth approximation would be to neglect $p$ and $q$ altogether w.r.t. $k$.  If we further restrict ourselves to small spin waves momenta $q<B/c$ (B is the optical bandwidth in units of frequency), then $q\ll p$ and certainly this approximation is valid. We will also assume a constant electronic density of states in the energy range $\Om=\pm \gamma$ where the integrand is appreciable and get a simple expression for the spin-spin correlations ({\em per unit volume}, and restoring missing $\hbar$'s)
\bea \label{spin-spin-2nd-order-3}
&& \la S_z(q,t)S_z(q',t')\ra^{(2)}=\delta_{q,-q'}\frac{\pi |d|^2 \rho_{el}\sum_p N_p}{\hbar\gamma_s}e^{-\gamma_s|t-t'|}\eea
where $\rho_{el}(\e_0)=\sqrt{2m^3\e}/\hbar^3|_{\e=\e_0}$ is the electron-hole density of states per unit volume in the bulk, taken at the central energy of excited electrons $\e_0=\hbar\om_0-E_g$, with $m$ the reduced e-h effective mass, and $\e$ the excess kinetic energy of the e-h pair above the gap.
Denoting $\la (S_z^2)_{q,w}\ra=F.T.[\la S_z(q,t)S_z(q',t')\ra]$ we obtain
\be
\label{spin-spin-2nd-order-4}
\la (S_z^2)_{q,\om} \ra^{(2)}=\frac{2\pi |d|^2 \rho_{el}\sum_p N_p}{\hbar(\om^2+\gamma_s^2)} =\frac{2\pi |d|^2 \rho_{el}\int d\om' \rho_{opt}(\om')N(\om')}{\hbar(\om^2+\gamma_s^2)}\ee
where $N(\om')\equiv N_q|_{\om'=cq}$ is the photon distribution function in frequency space, and $\rho_{opt}(\om)=\om^2/\pi^2c^3$ is the free space optical density of states per unit volume\footnote{Note that we redefine $d\rightarrow d\sqrt{V}$ to absorb the volume coming from the $k$-summation, so now it does not scale with $volume^{-1/2}$ anymore, and it has units of $energy\times volume^{1/2}$. Also note that while $\rho_{el}$ has units of $1/volume\times energy$, $\rho_{opt}$ has units of $1/volume\times frequency$.}. The contribution for long wavelengths is thus a Lorentzian background whose width is determined by the spin relaxation time.

Turning now to the fourth order contribution, we observe that in certain physical situations it (\ref{spin-spin-4th-1}) can be simplified considerably. This happens if the excess energy of the light above the electron-hole gap ($\hbar\om_0-E_g$) is large compared to the exciton binding energy. In this regime, using the correlations (\ref{four-fields-correlations}), and assuming that $k\gg p,q$ we get for the only important contribution the following approximate simple expression
\bea \label{spin-spin-4th-3}
&& \la S_z(q,t)S_z(q',t') \ra^{(4)}=\delta_{q,-q'}\frac{2\pi^2\rho_{el}^2|d|^4 e^{-i\om_q(t-t')}}{\hbar^2(\om_q^2+\gamma_s^2)}\sum_{p}\left[\right. \\ \nonumber
&& N_{p-q,+}N_{p,+}+N_{p,+}N_{p+q,+}+N_{p-q,-}N_{p,-}+N_{p,-}N_{p+q,-}+ \\ \nonumber
&& M^{(1)*}_{p-q,+}M^{(1)}_{p,+}+M^{(1)*}_{p,+}M^{(1)}_{p+q,+}+
M^{(1)*}_{p-q,-}M^{(1)}_{p,-}+M^{(1)*}_{p,-}M^{(1)}_{p+q,-}- \\ \nonumber
&& \left. -M^{(2)*}_{p-q,+}M^{(2)}_{p,+}-M^{(2)*}_{p-q,-}M^{(2)}_{p,-}-
M^{(2)*}_{p,+}M^{(2)}_{p+q,+}-M^{(2)*}_{p,-}M^{(2)}_{p+q,-}\right]
\eea
where $\rho_{el}$ is the electron-hole joint density of states at $\dek=\hbar\om_0$. Since the field is unpolarized $N_{+}=N_{-}$, and assuming again for simplicity the symmetric case $M_{q+}^{(\alpha)}=M_{q-}^{(\alpha)}=M_q^{(\alpha)}$, to which we always refer from this point on. Therefore we can omit the polarization indices
\bea \label{spin-spin-4th-4}
&& \la S_z(q,t)S_z(q',t') \ra^{(4)}=\delta_{q,-q'}\frac{4\pi^2\rho_{el}^2|d|^4 e^{-i\om_q(t-t')}}{\hbar^2(\om_q^2+\gamma_s^2)}
\sum_{p}\left[ N_{p-q}N_{p}+N_{p}N_{p+q}+\right. \\ \nonumber
&& \left. M^{(1)*}_{p-q}M^{(1)}_{p}+M^{(1)*}_{p}M^{(1)}_{p+q}-M^{(2)*}_{p-q}M^{(2)}_{p}-M^{(2)*}_{p}M^{(2)}_{p+q}\right].
\eea
The restriction $q'=-q$ which reflects translational invariance leads to $\la S_z(q,t)S_z(q',t') \ra=\la S_z(q,t')S_z(q',t) \ra^*$. The r.h.s. of Eq. (\ref{spin-spin-4th-4}) indeed obeys this if the summation over $p$ is real, which can be shown explicitly by using the symmetry property $M_p=M_{2p_0-p}$ of the squeezing correlation.

\begin{figure}\centering
\includegraphics[scale=0.7]{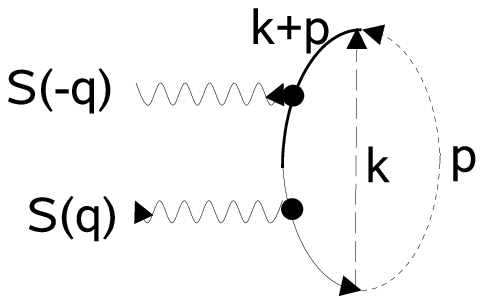}
\caption{Diagram of the second order contribution to the spin-spin correlations. The electron in the valence band (dashed line) is excited into the conduction band (solid line) by the photon correlation (dotted line). The only possible dynamics between the times $t,t'$ in this order of perturbation theory is due to the phenomenological spin relaxation ($\tau_s$).}
\label{diagram-second-order}\end{figure}

It is useful at this point to compare the decay and oscillatory time dependence of Eq. (\ref{spin-spin-2nd-order-3}) and Eq. (\ref{spin-spin-4th-3}) respectively. The second order contribution (\ref{spin-spin-2nd-order-3}) describes part of the spin fluctuations which is only due to the effect of the environment. The spin remains temporally correlated over a time scale $\tau_s$ that it takes for the relaxation processes to be effective. As can be seen from Fig. \ref{diagram-second-order}, the two orders of the interaction with the light are not sufficient to affect the dynamics of the correlation function $\la S_z(q,t)S_z(q',t') \ra$.
The contribution of the fourth order processes is physically different. Here the spin correlation dynamics reflects directly the correlations of the polarization of the light. This can be understood also diagrammatically from Fig. \ref{diagrams-sketch}. In the fourth order contribution there is no electronic propagator temporally connecting $S_z(q,t)$ and $S_z(q',t')$ (see Fig. \ref{diagrams-sketch}). Only the correlation functions of the light are connecting the two electronic diagrams, and the dynamics of the light is affecting the two times dynamics. We see that the spin relaxation $\gamma_s$ does affect the weight of each spectral component $S_z(q,t)$. Fast spectral components $\om_q\gg \gamma_s$ are independent of $\gamma_s$ and their weight is  $\propto \om_q^{-2}$, i.e. fast oscillations do not have enough time to be affected by the spin relaxation. In contrast, slow spectral components $\om_q\ll \gamma_s$ are building up slow enough to be balanced by $\gamma_s$, and their weight is approximately $\propto \gamma_s^{-2}$. In particular, without the spin relaxation in the model we would have a divergence of the $q=0$ component, which corresponds to an unbalanced accumulation of spin-correlated charges in the conduction band due to the double absorption process. Note that there are also other diagrams of the fourth order where $\gamma_s$ can affect the time dependence, however they have a much smaller contribution due to electronic phase space considerations (see discussion in appendix \ref{sec-1x3} and elsewhere\cite{paper2}).

The spectral function looks like
\bea \label{spin-spin-4th-5}
&& \la (S_z^2)_{q,\om} \ra^{(4)}=\frac{4\pi^2\rho_{el}^2|d|^4 \delta(\om-c q)}{\hbar^2(\om^2+\gamma_s^2)}
\sum_{p}\left[ N_{p-q}N_{p}+N_{p}N_{p+q}+\right. \\ \nonumber
&& \left. M^{(1)*}_{p-q}M^{(1)}_{p}+M^{(1)*}_{p}M^{(1)}_{p+q}-M^{(2)*}_{p-q}M^{(2)}_{p}-M^{(2)*}_{p}M^{(2)}_{p+q}\right].
\eea
The $\delta$-function reflects the dispersion of the light. The $p$ summation can be transformed to a frequency integral (so now the l.h.s. is {\em per unit volume})
\bea \label{spin-spin-4th-6}
&& \la (S_z^2)_{q,\om} \ra^{(4)}=\frac{4\pi^2\rho_{el}^2|d|^4 \delta(\om-\om_q)}{\hbar^2(\om^2+\gamma_s^2)}
\int d\om' \rho_{opt}(\om')\left[ N(\om'-\om)N(\om')+N(\om')N(\om'+\om)+\right. \\ \nonumber
&& \left. M^{(1)*}(\om'-\om)M^{(1)}(\om')+M^{(1)*}(\om')M^{(1)}(\om'+\om)-M^{(2)*}(\om'-\om)M^{(2)}(\om')-
M^{(2)*}(\om')M^{(2)}(\om'+\om)\right].
\eea
We see that as the phases of $M^{(1,2)}$ are modulated as a function of frequency, the spectrum of the driven spin fluctuations is also modulated. The spin structure factor in these situations is in fact \textit{coherently controlled} using the phases of the $\om$-dependent squeezing.
Combining the second (\ref{spin-spin-2nd-order-4}) and fourth (\ref{spin-spin-4th-6}) order contributions, defining
\bea \label{convolution function}
&& C(\om)=\int d\om' \left[ N(\om'-\om)N(\om')+N(\om')N(\om'+\om)+\right. \\ \nonumber
&&  M^{(1)*}(\om'-\om)M^{(1)}(\om')+M^{(1)*}(\om')M^{(1)}(\om'+\om)- \\ \nonumber
&& \left. -M^{(2)*}(\om'-\om)M^{(2)}(\om')-M^{(2)*}(\om')M^{(2)}(\om'+\om)\right]
\eea
and taking the optical density of states as approximately constant, given by $\rho_{opt}=\rho_{opt}(\om_0)$ from the middle of the spectrum, we obtain
\bea \label{spin-spin-combined-1}
&& \la (S_z^2)_{q,\om} \ra^{(2+4)}=\frac{2\pi |d|^2 \rho_{el}\rho_{opt}}{\hbar(\om^2+\gamma_s^2)} \left[ \int d\om' N(\om') +4\pi^2\frac{\rho_{el}|d|^2}{\hbar} \delta(\om-cq) C(\om) \right]\eea
Note that this result is valid for long wavelengths ($q \ll B/c$), which is also the range where the spin structure factor is largest.

Let us now choose as direction of light propagation $\hat{z}$ and integrate over $q_z$, setting $q_x=q_y=0$. Since our approximations are only valid for small $q$, we will use a cutoff $B/c$ on the $q_z$ integration, which mean that the result is the $\om$-spectrum spatially averaged over the coherence length $c/B$ in $z$-direction and the $x-y$ plane. Physically, this is motivated by having in mind a Faraday probe beam passing through the sample and measuring the spin fluctuations. For small frequencies $\om\ll B$ this integration will smear-out the $\delta$-function singularity. We assume that the sample is much smaller compared to the coherence length of the light $l_c$, so the averaging is effectively done over the whole sample. 
This integration gives the following spin spectral density
\be \label{spin-spin-combined-2}
\la (S_z^2)_{\om} \ra^{(2+4)}=\int_{-B/c}^{B/c} dq_z \la (S_z^2)_{q,w} \ra^{(2+4)}|_{_{q_x=q_y=0}}=\frac{4\pi |d|^2 \rho_{el}\rho_{opt}B}{\hbar c(\om^2+\gamma_s^2)} \left[ \int d\om' N(\om') +\frac{4\pi^2\rho_{el}|d|^2}{\hbar B}C(\om) \right].\ee

In order to estimate the strength of the coherent effects compared to the background of second order fluctuations we need to calculate the ratio of the 4th order to the 2nd order \be \mathcal{R}(\om)=\frac{4\pi^2 \rho_{el} d^2}{\hbar B}\frac{C(\om)}{\int d\om' N(\om')} \ee
which at $\om=0$ can be approximately estimated for strong classical squeezing (e.g. $M^{(1)}=N$) to give $16\pi^2 \rho_{el} d^2 \bar{N}/\hbar B$ where $\bar{N}$ is the average photon occupation inside the optical bandwidth.
Since this is a perturbative calculation, we expect the result, Eq. (\ref{spin-spin-combined-2}) to be valid as long as the second term is much smaller than the first one. This will generally put a restriction on the intensity, or photon occupation function $N(\om)$, for a given optical bandwidth $B$. For an optical excitation which generates conduction band electrons with kinetic energy of about $30meV$, and an optical energy of approximately $1.5eV$, we can estimate that for an optical bandwidth $\hbar B\approx 10meV$ the average photon occupation should be $10^{-4}-10^{-5}$ for the second term to be $1/10$ of the first one ($\mathcal{R}(\om=0)=0.1$). Usually if one uses correlated photons from down-converted light, the spectrum is very wide in the non-degenerate case, and therefore the average photon number per mode ($\bar{N}$) is very small. For example as discussed in Wang et al.\cite{wang} for type-I down converted light, a counting rate of $10^4$ photons/sec was reported over a bandwidth of $10^{12}Hz$, which corresponds to $\bar{N}=10^{-8}$. In more recent experiments, Bowen et al.\cite{bachor}, Heersink et al. \cite{leuchs-2} and Marquardt et al. \cite{Marquardt} report of higher intensities for polarization-squeezed light. 

It is interesting to note that according to the above estimates there appears to exist an experimentally accessible regime where the perturbative approximation breaks down ($N\geq \hbar B/16\pi^2 \rho_{el} d^2$). In this regime higher orders in the light-matter interaction become important,  leading to a strong non-linear response to the driving field. It is an intriguing direction for future research, especially since strong correlated light sources are becoming increasingly available \cite{leuchs-1,leuchs-2}. Another interesting direction to explore is related to effects of non-classical squeezed light $|M|>N$ on the spectrum, which we showed to have a unique effect on the static spin correlations \cite{paper2}.


\section{Examples of phase modulations}\label{sec-results}

The quantity $C(\om)$ appearing in Eq. (\ref{spin-spin-combined-2}) is sensitive to the squeezing phases, i.e. the phases of $M(\om)$, cf. Eq. (\ref{convolution function}). In the general case $arg(M(\om))$ can be expanded in a power series so initially it makes sense to focus on the linear and quadratic dependence. First let us rewrite the spectral spin density (\ref{spin-spin-combined-2}) as
\be \label{spin-spin-combined-3}
\int_0^{B/c} dq_z \la (S_z^2)_{q,w} \ra^{(2+4)}|_{_{q_x=q_y=0}}=\alpha \left(\tilde{\om}^2+(\frac{\gamma_s}{B})^2\right)^{-1} \left[ 1 +\eta \,\mathcal{C}(\tilde{\om}) \right]\ee
where $\tilde{\om}=\om/B$, $\alpha=\frac{2\pi \rho_{opt}d^2 \rho_{el}}{cB}\int d\om' N(\om')$, $\eta=4\pi^2 \frac{\rho_{el}}{\rho_{opt}}\frac{\gamma_{rec}}{B}\bar{N}$, with the electron-hole recombination rate $\gamma_{rec}=\frac{2\pi}{3}\rho_{opt}d^2/\hbar^2$, the average photon occupation $\bar{N}$, and  $\mathcal{C}(\tilde{\om})=B^{-1}\bar{N}^{-2}C(B\tilde{\om})$.

\begin{figure}\centering
\includegraphics[scale=0.8]{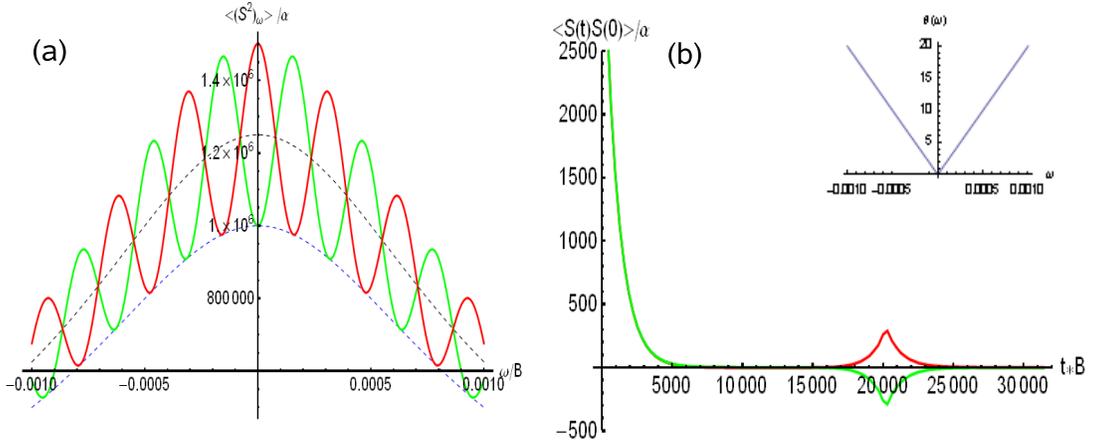}
\caption{(a) $\om$-dependence of the spin structure factor for a linear phase modulation of the squeezing phase of the field (insert: the dependence of the phase on the frequency). The figure depicts the 2nd order contribution (dashed blue), and phase sensitive 4th order contributions for same-polarization (red) and opposite-polarization squeezing (green), respectively, compared to no squeezing (dashed black). (b) In the time domain we clearly see that the phase modulation parameter $T=20000s$ determines appearance time of the second correlation peak. The correlation between polarizations determine whether this peak will be positive (red) or negative (green).}
\label{linear-phase-mod}\end{figure}

In the linear case we take $\theta(\om-\om_0)=T|\om-\om_0|$ with a phase modulation parameter $T=2\times 10^4 sec$, small average photon number $\bar{N}=10^{-4}$, optical bandwidth $\hbar B=20 meV$, average electron kinetic energy $\e_k=30meV$, and spin relaxation time $\tau_s=0.2ns$. The photon correlation functions were taken to be Gaussian $N(\om)=exp[-((\om-\om_0)/B)^2]$ and $M(\om)=N(\om)exp[iT|\om-\om_0|]$. One possible experimental realization of a phase modulation of squeezed light is by using a pulse shaper \cite{dayan}. With these parameters such a pulse shaper has to turn the squeezing phase a full cycle of $2\pi$ every 1.6GHz.
In Fig. \ref{linear-phase-mod}(a) we draw the r.h.s. of (\ref{spin-spin-combined-3}) divided by $\alpha$ as a function of $\tilde{\om}$. We see that the spectrum develops as additional oscillatory structure with a frequency $T^{-1}$, which depends on the type of squeezing (same or opposite polarization). The linear phase modulation in frequency space can be thought of as a time-shift in the optical field in time domain, Fig. \ref{linear-phase-mod}(b). This is somewhat similar to a retardation of part of the random polarization signal, leading to the second rise of the spin correlation exactly after that time ($T$). This analogy is however not exact since the frequency shift is a symmetric function \cite{Itay}.

\begin{figure}\centering
\includegraphics[scale=0.8]{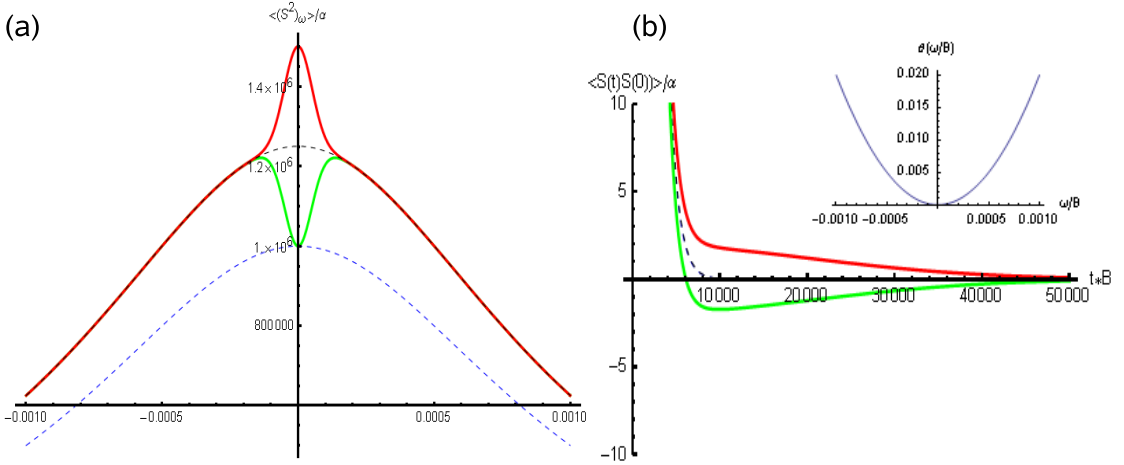}
\caption{(a) $\om$-dependence of the spin structure factor for a quadratic phase modulation of the squeezing phase of the field (insert: the dependence of the phase on the frequency). The figure depicts the 2nd order contribution (blue), and phase sensitive 4th order contributions for same-polarization (red) and opposite-polarization squeezing, respectively, compared to no squeezing (green). (b) In time domain, in contrast to the linear case, the chirped phase modulation endows the spin correlation with a long tail of correlation determined by $T$ rather than by $\tau_s$, and starting immediately after the time $\tau_s$. Again the polarization correlation determines the sign of the long range temporal correlation.}
\label{quad-phase-mod}\end{figure}

In the quadratic case (see Fig. \ref{quad-phase-mod}), we take $\theta(\om-\om_0)=T(\om-\om_0)^2$ with a phase modulation parameter $T=2\times 10^4 sec^2$, small average photon number $\bar{N}=10^{-4}$, optical bandwidth $\hbar B=20 meV$, average electron kinetic energy $\e_k=30meV$. The photon correlation functions were taken to be again Gaussian with the squeezing $M(\om)=N(\om)exp[iT(\om-\om_0)^2]$. In this case it is clear that the effect of 'chirping' in $\om$-space, leads through the non-linear formula Eq. (\ref{spin-spin-combined-2}) to an enhancement (reduction) of the correlations around the DC component of the spin correlations when the light is endowed with same (opposite) polarization correlations. In the time domain we see that there are two time scales, $\gamma_s^{-1}$ and $T$ that govern the behavior of the spin correlations.
For opposite polarization correlations we observe a qualitative difference: on short time scales $t< \gamma_s^{-1}$ the correlation is positive but for longer times $t>T$ it becomes negative. This change only happens in the opposite polarizations case when the squeezing phase is modulated on a scale much smaller than the spin relaxation rate ($T^{-1}\ll \gamma_s$).

With the experimental parameters estimated above, $\eta\approx 0.1$, and $\alpha\approx 10^9 sec/m^4$.
To estimate the strength of the total spin fluctuations we can look at the prefactor $\alpha/(\tilde{\om}^2+(\gamma_s/B)^2)$ for $\tilde{\om}=0$ or at the integrated power, depending on what kind of experiment is performed. For the first case we get that the prefactor scales like $\bar{N}B^2$, while $\eta\propto \bar{N}/B$, so the overall strength of the spin fluctuations can be increased while keeping $\eta$ small. Since light with frequencies above the e-h gap is strongly absorbed with a typical attenuation coefficient of the order of $10^{-3}cm^{-1}$ for GaAs, it is preferable to assume a thin slab, for example with a thickness of $5\mu m$ and a beam of light with an area of $1mm^2$. For this geometry and the other parameters given above, the strength of the spin fluctuations around $\om=0$ would be between $10-1000$ in units of $\hbar^2$, with the range reflecting an uncertainty of several parameters. This is the strength of the combined contribution of the second and fourth order, so the signal (the fourth order) is estimated from $\eta$ to be around $10\%$. Even though the total spin is very small for a macroscopic sample, we note that recently even a {\em single spin} has been measured with the Faraday rotation technique \cite{imamoglu}.


%

\appendix
\section{Optical transition matrix elements}\label{app-A}

We summarize here for convenience the calculation of the dipole interaction matrix elements $d_{k,p}^{s}$ that appear in Eq. (\ref{H-LM})) for optical transitions in direct-gap semiconductors. In the spherical version of the Luttinger-Kohn model\cite{Luttinger-1} for degenerate valence bands we can choose the angular momentum quantization axis $\hat{z}$ to be parallel to the optical beam $\hat{n}$. This model takes into account the spin-orbit interaction in the valence band, which lead to mixing of the {\em heavy-hole} (HH) and {\em light-hole} (LH) states ($j=3/2$). For the fourth order correlation processes which we consider, the strongest contributions from $LH\rightarrow C$ transitions are still weaker by a factors of $1/3$. For this reason and for simplicity we consider only $HH \rightarrow C$ in our model.
Let us start with a given electronic wave vector $k$ near the $\Gamma$-point in the valence band. The wave function can be written as \cite{optical-orientation}
\be \psi^v_{k,m}=e^{ik\cdot r}\sum_{\mu} \chi_{_{m\mu}}(k)u^v_{\mu} \ee
where $\mu,m=\pm 3/2$ denote the HH eigenstates, $\chi_{_{m\mu}}(k)=\mathcal{D}^{(3/2)}_{\mu,m}(\varphi,\theta,\phi)$ are rotation matrices \cite{edmonds-book}($\theta,\varphi$ are polar angles of $\hat{k}$, $\phi$ a global phase), and $u^v_{\mu}$ are angular momentum eigenfunctions of $J_z$. The quantum number $m=\pm 3/2$ denotes the angular momentum projection in the direction of $\hat{k}$, and $\mu=\pm 3/2$ is the projection in the direction $z$. Using the usual notation for the wave functions of $s$ and $p$ orbitals,  $u^v_{3/2}=-\frac{1}{\sqrt{2}}(X+iY)\uparrow$ and $u^v_{-3/2}=\frac{1}{\sqrt{2}}(X-iY)\downarrow$. For the conduction band $\psi^c_{k,s}=e^{ik\cdot r} u_s$, where $u^c_{1/2}=S\uparrow$ and $u^c_{-1/2}=S\downarrow$. The electronic part of the dipole matrix elements are then given by
\be \la \psi^c_{k,s}|\vec{D}|\psi^v_{k',m}\ra=\delta_{k,k'}\sum_{\mu}\chi_{_{m\mu}}(k) \la u^c_s|\vec{D}|u^v_{\mu}\ra\ee
where $\vec{D}=e\vec{r}$ is the dipole moment operator. Let us consider the matrix elements for the transition into the state $k,s=1/2$
\bea
&& \la \psi^c_{k,1/2}|\vec{D}|\psi^v_{k',3/2}\ra= \\ \nonumber
&& =\delta_{k,k'}\left[ \chi_{_{+3/2,+3/2}}(k)\la S \uparrow | \vec{D} | -\frac{1}{\sqrt{2}}(X+iY) \uparrow \ra + \chi_{_{+3/2,-3/2}}(k)\la S\uparrow|\vec{D}|\frac{1}{\sqrt{2}}(X-iY)\downarrow \ra  \right]=  \delta_{k,k'} \chi_{_{+3/2,+3/2}}(k) \frac{-D}{\sqrt{2}} (\hat{x}+i\hat{y})  \\ \nonumber
&& \la \psi^c_{k,1/2}|\vec{D}|\psi^v_{k',-3/2}\ra= \\ \nonumber
&& =\delta_{k,k'}\left[ \chi_{_{-3/2,+3/2}}(k)\la S \uparrow | \vec{D} | -\frac{1}{\sqrt{2}}(X+iY) \uparrow \ra + \chi_{_{-3/2,-3/2}}(k)\la S\uparrow|\vec{D}|\frac{1}{\sqrt{2}}(X-iY)\downarrow \ra  \right]=\delta_{k,k'} \chi_{_{-3/2,+3/2}}(k) \frac{-D}{\sqrt{2}}(\hat{x}+i\hat{y})  \\ \nonumber
\eea
where we used the fact\cite{optical-orientation} that the only non-zero matrix elements are $\la S|D_x|X\ra=\la S|D_y|Y\ra=\la S|D_z|Z\ra=D$ due to the spatial symmetry of these integrals.
We see that due to spin orthogonality only one of the orbital angular component, $\mu=+3/2$ contributes to the transition into $+1/2$ state (a similar expression can be derived for $s=-1/2$). The full interaction matrix elements are given by $d_{k,k'-k}^{\lm,s,m}=\mathcal{E}_{\om_{k-k'}}\hat{\e}_{\lm}\cdot \la \vec{D}\ra_{k,k'}^{s,m}$, where $\lm=\pm$ denotes the helicity and $\mathcal{E}_{\om_{k-k'}}=\sqrt{\hbar \om_{k-k'}/2\varepsilon_0 V}$. We see that $\lm$ is determined by the conduction band spin, i.e. $s \Leftrightarrow\lm$, while for an arbitrary direction of $\hat{k}$ both $m=\pm 3/2$ contribute to the transition. We can therefore omit the index $\lm$ and summation over it, and continue with a simplified interaction Hamiltonian
\be \sum_{k,p}\sum_{s,m} d_{k,p}^{s,m} \cd_{k+p,s}b_{p,s}v_{k,m} + h.c. \ee
where we now use $s$ also for the helicity. By applying a transformation to the basis where  angular momentum projection is along $\hat{n}$ with new operators $\tilde{v}_{k,\mu}$
\be v_{k,m}=\chi_{_{m,+3/2}}(k)\tilde{v}_{k,+3/2}+\chi_{_{m,-3/2}}(k)\tilde{v}_{k,-3/2} \ee
the interaction appears as
\bea
&& \sum_{k,p}\sum_{s,m} d_{k,p}^{s,m} \cd_{k+p,s}b_{p,s}(\chi_{_{m,+3/2}}(k)\tilde{v}_{k,+3/2}+\chi_{_{m,-3/2}}(k)\tilde{v}_{k,-3/2}) + h.c. =\\ \nonumber
&& =\sum_{k,p}\sum_{s}  \cd_{k+p,s}b_{p,s}\left[ (d_{k,p}^{s,+3/2} \chi_{_{+3/2,+3/2}}(k)\tilde{v}_{k,+3/2}+d_{k,p}^{s,+3/2}\chi_{_{+3/2,-3/2}}(k)\tilde{v}_{k,-3/2})+ \right. \\  \nonumber
&& \left. +(d_{k,p}^{s,-3/2} \chi_{_{-3/2,+3/2}}(k)\tilde{v}_{k,+3/2}+d_{k,p}^{s,-3/2}\chi_{_{-3/2,-3/2}}(k)\tilde{v}_{k,-3/2})\right] + h.c.= \\ \nonumber
&&= \sum_{k,p}\sum_{s}  \cd_{k+p,s}b_{p,s}  \left[ (d_{k,p}^{s,+3/2}\chi_{_{+3/2,+3/2}}(k)+d_{k,p}^{s,-3/2}\chi_{_{-3/2,+3/2}}(k))\tilde{v}_{k,+3/2}+ \right. \\  \nonumber
&& \left. +(d_{k,p}^{s,+3/2}\chi_{_{+3/2,-3/2}}(k)+d_{k,p}^{s,-3/2}\chi_{_{-3/2,-3/2}}(k))\tilde{v}_{k,-3/2}\right] + h.c. \\ \nonumber
\eea
The rotation matrix element can be written as\cite{edmonds-book} $\mathcal{D}_{m',m}^{(j)}(\varphi,\theta,\phi)=e^{im'\varphi}d^{(j)}_{m',m}(\theta)e^{im\phi}$ (here $d$ is {\em not} the dipole matrix element).
By using the symmetry relation $d^{(j)}_{m',m}(\theta)=(-1)^{m'-m}d^{(j)}_{-m',-m}(\theta)$ and setting $\phi=0$, it is straightforward to show that the coefficient in front of $\tilde{v}_{k,+3/2}$ vanishes when $s=-1/2$ and the coefficient of $\tilde{v}_{k,-3/2}$ vanishes when $s=+1/2$. Therefore there is a one-to-one correspondence between the projection of angular momentum of the initial valence band state $\mu$ and the projection of the spin of the final conduction band state $s$. Renaming $\tilde{v}$ as $v$, we can rewrite the interaction term as
\bea
&& \sum_{k,p}\sum_{s}  \cd_{k+p,s}b_{p,s} (-s) D \mathcal{E}_{\om_p}(\chi^2_{s,s}(k)+\chi^2_{s,-s}(k)) v_{k,s} +h.c. \\ \nonumber
\eea
where a common index $s=\pm 1\Leftrightarrow \pm 1/2 \Leftrightarrow \pm 3/2$ is used for angular momentum, spin, and helicity.
In cases where the angular dependence does not lead to other consequences other than some numerical prefactor (in cases where an isotropic approximation for all the other parts in the integrand is reasonable), we just omitted it and assume that $D$ absorbs the prefactor.


\section{Phenomenological level widths}\label{app-B}

The fourth order contribution $\la S(t)S(t')\ra_{s.s.}^{(4)}$ which we calculate here should match our previously calculated injection rate of static correlations \cite{paper2} in the limit $t'\rightarrow t$. Indeed it can be shown that they give the same result if in the previous calculation a finite level width $\tau_s^{-1}$ is incorporated for the finite state propagator, and the intermediate state level width is taken to zero.
\begin{figure}\centering
\includegraphics[scale=0.8]{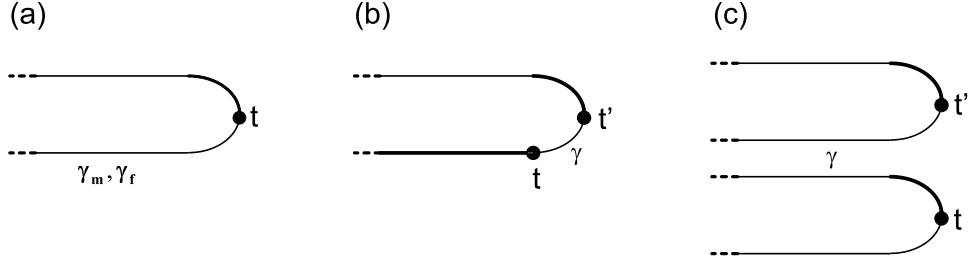}
\caption{(Color online) A schematic for the time evolution for (a) direct evaluation of the static correlations via second order perturbation theory (b) the two-times spin fluctuation function with a dissipative part shown in grey, and (c) the evaluation of the same correlation with equations of motion method.}
\label{unitarity-fig}\end{figure}
To explain the source of this problem we need to compare the two approaches. Previously we calculated the static correlation ($t=t'$) directly, using a second order perturbation expansion of the wave function.
In interaction picture the static correlation can be written as (see Fig. \ref{unitarity-fig}(a))
\be \label{unitarity-1}\la ~\la\psi_{el}|U^{\dagger}(t,-\infty)S(t)_0^2 U(t,-\infty)|\psi_{el}\ra ~\ra_{field}\ee
while here we calculate the two-times correlation, which is defined as (see Fig. \ref{unitarity-fig}(b))
\be \label{unitarity-2} \la ~\la\psi_{el}|U^{\dagger}(t,-\infty)S(t)_0 U(t,t')S(t')_0 U(t',-\infty)|\psi_{el}\ra ~\ra_{field}\ee
where $S(t)_0$ denote operators in the interaction picture.
Since a direct evaluation of (\ref{unitarity-2}) is very difficult for fourth order processes, we use the equations of motions for $S(t),S(t')$, which means actually that we split the intermediate time evolution  into a product which gives (see Fig. \ref{unitarity-fig}(c))
\bea \label{unitarity-3} && \la ~\la\psi_{el}|\tilde{U}^{\dagger}(t,-\infty)S(t)_0 \tilde{U}(t,-\infty)\times \\ \nonumber
&& \tilde{U}^{\dagger}(t',-\infty)S(t')_0 \tilde{U}(t',-\infty)|\psi_{el}\ra ~\ra_{field}\eea
which is in principle equivalent to (\ref{unitarity-2}) due to unitarity.
The difficult issue is how to account also for the relaxation processes in such calculation in the simplest way while retaining consistency. In our calculations we added them explicitly as level broadenings ($\gamma$). However when the relaxation is added to the time evolution it is no longer possible to claim that unitarity applies
\be \label{unitarity-4} \tilde{U}(t,-\infty;\gamma) \tilde{U}^{\dagger}(t',-\infty;\gamma)\neq U(t,t';\gamma)\ee
and therefore in principle different approaches can give different results. Specifically for the limit $t\rightarrow t'$ the evolution depicted in Fig. \ref{unitarity-fig}(a) might not give the same result as going along ($t \rightarrow -\infty \rightarrow t$) in Fig. \ref{unitarity-fig}(c).
In principle this shows that adding dissipations phenomenologically in higher order perturbation theory is very tricky and it would be better to have a microscopic model instead. We have shown\cite{mythesis} another approach with Keldysh diagrammatic formalism that is promising to lead in the future to a derivation in which the dissipation processes will be derived from a microscopic Hamiltonian, and will not suffer from such ambiguities.

\section{Calculation of the $\la S_z^{(1)}(q,t)S_z^{(3)}(q',t')\ra$ contribution}\label{sec-1x3}

We begin by writing explicitly the second contribution to the spin wave operator
\bea \label{sz-second-order-explicit}
&& S_z(q,t)^{(2)}=  -|d|^2 e^{-\gamma_s t}\int_{t_0}^t dt_1 e^{-\gamma t_1}e^{\gamma_s t_1}\int_{t_0}^{t_1}dt_2   e^{\gamma t_2}\sum_{k,p,p'}\left[  \right.\\ \nonumber
&& e^{i(\D\e_{k-q,k-p}-\om_p)t_1}e^{-i(\D\e_{k-q,k-p}-\om_{p'})t_2} b^*_{p'\ua}\left( n^c_{k-q,k-p+p',\ua}(t_2)_0-n^v_{k-q-p',k-p,\ua}(t_2)_0\right) b_{p\ua}+  \\ \nonumber
&& +e^{-i(\D\e_{k,k-p-q}-\om_p)t_1}e^{i(\D\e_{k,k-p-q}-\om_{p'})t_2} b^*_{p\ua} \left(n^c_{k-p-q+p',k,\ua}(t_2)_0-n^v_{k-p-q,k-p',\ua}(t_2)_0\right)b_{p'\ua}- \\ \nonumber
&& -e^{i(\D\e_{k-q,k-p}-\om_p)t_1}e^{-i(\D\e_{k-q,k-p}-\om_{p'})t_2} b^*_{p'\da}\left( n^c_{k-q,k-p+p',\da}(t_2)_0-n^v_{k-q-p',k-p,\da}(t_2)_0\right) b_{p\da}-   \\ \nonumber
&& \left. - e^{-i(\D\e_{k,k-p-q}-\om_p)t_1}e^{i(\D\e_{k,k-p-q}-\om_{p'})t_2} b^*_{p\da} \left(n^c_{k-p-q+p',k,\da}(t_2)_0-n^v_{k-p-q,k-p',\da}(t_2)_0\right)b_{p'\da}\right]
\eea
In order to derive the third order contribution $S_z(q,t)^{(3)}$ it is necessary to formally solve the equations of motion for the various operators that appear in the second order contribution (\ref{sz-second-order-explicit}).
We start by writing integral expressions for $n^{c,v}_{k,k'}$, again employing the basic equations of motion (\ref{eoms-basic}), and repeat the approximations of static spin waves, by neglecting the free evolution of $n_{k,k'}$ terms
\bea \nonumber
&& n^c_{k-q,k-p+p',\s}(t_2)= \\ \nonumber
&& =e^{-\gamma t_2}\int_{t_0}^{t_2}dt_3 e^{\gamma t_3}i\sum_{p''}\left[ -d\pd_{k-p+p'-p'',k-q,\s}(t_3)_0b_{p''\s}(t_3)_0+
d^*b^*_{p''\s}(t_3)_0P_{k-q-p'',k-p+p',\s}(t_3)_0\right]= \\ \nonumber
&&=\sum_{p''}e^{-\gamma t_2}\int_{t_0}^{t_2}dt_3 e^{\gamma t_3}i\left[-e^{i(\D\e_{k-q,k-p+p'-p''}-\om_{p''})t_3}
d\pd_{k-p+p'-p'',k-q,\s}b_{p''\s}+ \right. \\ \nonumber
&& \left. +e^{-i(\D\e_{k-p+p',k-q-p''}-\om_{p''})t_3}d^*b^*_{p''\s}P_{k-q-p'',k-p+p',\s}\right]= \\ \nonumber
&& =\sum_{p''}\left[ \frac{-e^{i(\D\e_{k-q,k-p+p'-p''}-\om_{p''})t_2}d\pd_{k-p+p'-p'',k-q,\s}b_{p''\s}}
{\gamma+i(\D\e_{k-q,k-p+p'-p''}-\om_{p''})} + \right. \\ \nonumber
&& + \left. \frac{e^{-i(\D\e_{k-p+p',k-q-p''}-\om_{p''})t_2}d^*b^*_{p''\s}P_{k-q-p'',k-p+p',\s}}
{\gamma-i(\D\e_{k-p+p',k-q-p''}-\om_{p''})} \right]\\
\eea
\bea \nonumber
&& n^c_{k-p+p'-q,k,\s}(t_2)= \\ \nonumber
&& =e^{-\gamma t_2}\int_{t_0}^{t_2}dt_3 e^{\gamma t_3}i\sum_{p''}\left[ -d\pd_{k-p'',k-p+p'-q,\s}(t_3)_0b_{p''\s}(t_3)_0+
d^*b^*_{p''\s}(t_3)_0P_{k-p+p'-p''-q,k,\s}(t_3)_0\right]= \\ \nonumber
&&=\sum_{p''}e^{-\gamma t_2}\int_{t_0}^{t_2}dt_3 e^{\gamma t_3}i\left[-e^{i(\D\e_{k-p+p'-q,k-p''}-\om_{p''})t_3}
d\pd_{k-p'',k-p+p'-q,\s}b_{p''\s}+ \right. \\ \nonumber
&& \left. +e^{-i(\D\e_{k,k-p+p'-p''-q}-\om_{p''})t_3}d^*b^*_{p''\s}P_{k-p+p'-p''-q,k,\s}\right]= \\ \nonumber
&& =\sum_{p''}\left[ \frac{-e^{i(\D\e_{k-p+p'-q,k-p''}-\om_{p''})t_2}d\pd_{k-p'',k-p+p'-q,\s}b_{p''\s}}
{\gamma+i(\D\e_{k-p+p'-q,k-p''}-\om_{p''})} + \right. \\ \nonumber
&& + \left. \frac{e^{-i(\D\e_{k,k-p+p'-p''-q}-\om_{p''})t_2}d^*b^*_{p''\s}P_{k-p+p'-p''-q,k,\s}}
{\gamma-i(\D\e_{k,k-p+p'-p''-q}-\om_{p''})} \right]\\
\eea
\bea \nonumber
&& n^v_{k-p'-q,k-p,\s}(t_2)= \\ \nonumber
&& =e^{-\gamma t_2}\int_{t_0}^{t_2}dt_3 e^{\gamma t_3}i\sum_{p''}\left[ d\pd_{k-p,k-p'-q+p'',\s}(t_3)_0b_{p''\s}(t_3)_0-
d^*b^*_{p''\s}(t_3)_0P_{k-p'-q,k-p+p'',\s}(t_3)_0\right]= \\ \nonumber
&&=\sum_{p''}e^{-\gamma t_2}\int_{t_0}^{t_2}dt_3 e^{\gamma t_3}i\left[e^{i(\D\e_{k-p'-q+p'',k-p}-\om_{p''})t_3}
d\pd_{k-p,k-p'-q+p'',\s} b_{p''\s}- \right. \\ \nonumber
&& \left. -e^{-i(\D\e_{k-p+p'',k-p'-q}-\om_{p''})t_3}d^*b^*_{p''\s}P_{k-p'-q,k-p+p'',\s}\right]= \\ \nonumber
&& =\sum_{p''}\left[ \frac{e^{i(\D\e_{k-p'-q+p'',k-p}-\om_{p''})t_2}d\pd_{k-p,k-p'-q+p'',\s} b_{p''\s}}
{\gamma+i(\D\e_{k-p'-q+p'',k-p}-\om_{p''})} - \right. \\ \nonumber
&& - \left. \frac{e^{-i(\D\e_{k-p+p'',k-p'-q}-\om_{p''})t_2}d^*b^*_{p''\s}P_{k-p'-q,k-p+p'',\s}}
{\gamma-i(\D\e_{k-p+p'',k-p'-q}-\om_{p''})} \right]\\
\eea
\bea \nonumber
&& n^v_{k-p-q,k-p',\s}(t_2)= \\ \nonumber
&& =e^{-\gamma t_2}\int_{t_0}^{t_2}dt_3 e^{\gamma t_3}i\sum_{p''}\left[ d\pd_{k-p',k-p+p''-q,\s}(t_3)_0b_{p''\s}(t_3)_0-
d^*b^*_{p''\s}(t_3)_0P_{k-p-q,k-p'+p'',\s}(t_3)_0\right]= \\ \nonumber
&&=\sum_{p''}e^{-\gamma t_2}\int_{t_0}^{t_2}dt_3 e^{\gamma t_3}i\left[e^{i(\D\e_{k-p+p''-q,k-p'}-\om_{p''})t_3}
d\pd_{k-p',k-p+p''-q,\s} b_{p''\s}- \right. \\ \nonumber
&& \left. -e^{-i(\D\e_{k-p'+p'',k-p-q}-\om_{p''})t_3}d^*b^*_{p''\s}P_{k-p-q,k-p'+p'',\s}\right]= \\ \nonumber
&& =\sum_{p''}\left[ \frac{e^{i(\D\e_{k-p+p''-q,k-p'}-\om_{p''})t_2}d\pd_{k-p',k-p+p''-q,\s} b_{p''\s}}
{\gamma+i(\D\e_{k-p+p''-q,k-p'}-\om_{p''})} - \right. \\ \nonumber
&& - \left. \frac{e^{-i(\D\e_{k-p'+p'',k-p-q}-\om_{p''})t_2}d^*b^*_{p''\s}P_{k-p-q,k-p'+p'',\s}}
{\gamma-i(\D\e_{k-p'+p'',k-p-q}-\om_{p''})} \right]\\
\eea

When now substituting $n^{v,c}_{k,k'}$ in the expression for $S_z^{(2)}(q,t)$, we only need the $\propto \pd$ terms, because in the correlation $1\times 3$ only averages of the form $\la P \pd\ra$ are contributing for zero temperature, leading to
\bea
&& S_z(q,t)^{(3)}=  -|d|^2 e^{-\gamma_s t}\int_{t_0}^t dt_1 e^{-\gamma t_1}e^{\gamma_s t_1}\int_{t_0}^{t_1}dt_2   e^{\gamma t_2}\sum_{k,p,p',p''}\left[  \right.\\ \nonumber
&& e^{i(\D\e_{k-p,k-q}-\om_p)t_1}e^{-i(\D\e_{k-p,k-q}-\om_{p'})t_2} b^*_{p'\ua}\left( \frac{-e^{i(\D\e_{k-p+p',k-q-p''}-\om_{p''})t_2}d\pd_{k-p+p'-p'',k-q,\ua}b_{p''\ua}}
{\gamma+i(\D\e_{k-p+p',k-q-p''}-\om_{p''})}- \right.\\ \nonumber
&& \left. -\frac{e^{i(\D\e_{k-p+p'',k-p'-q}-\om_{p''})t_2}d\pd_{k-p,k-p'-q+p'',\ua} b_{p''\ua}}
{\gamma+i(\D\e_{k-p+p'',k-p'-q}-\om_{p''})}\right) b_{p\ua}+  \\ \nonumber
&& +e^{-i(\D\e_{k-p-q,k}-\om_p)t_1}e^{i(\D\e_{k-p-q,k}-\om_{p'})t_2} b^*_{p\ua} \left(\frac{-e^{i(\D\e_{k-p+p'-q,k-p''}-\om_{p''})t_2}d\pd_{k-p'',k-p+p'-q,\ua}b_{p''\ua}}
{\gamma+i(\D\e_{k-p+p'-q,k-p''}-\om_{p''})}- \right. \\ \nonumber
&& \left. - \frac{e^{i(\D\e_{k-p+p''-q,k-p'}-\om_{p''})t_2}d\pd_{k-p',k-p+p''-q,\ua} b_{p''\ua}}
{\gamma+i(\D\e_{k-p+p''-q,k-p'}-\om_{p''})}\right)b_{p'\ua}- \\ \nonumber
&& -e^{i(\D\e_{k-p,k-q}-\om_p)t_1}e^{-i(\D\e_{k-p,k-q}-\om_{p'})t_2} b^*_{p'\da}\left( \frac{-e^{i(\D\e_{k-p+p',k-q-p''}-\om_{p''})t_2}d\pd_{k-p+p'-p'',k-q,\da}b_{p''\da}}
{\gamma+i(\D\e_{k-p+p',k-q-p''}-\om_{p''})}- \right.\\ \nonumber
&& \left. -\frac{e^{i(\D\e_{k-p+p'',k-p'-q}-\om_{p''})t_2}d\pd_{k-p,k-p'-q+p'',\da} b_{p''\da}}
{\gamma+i(\D\e_{k-p+p'',k-p'-q}-\om_{p''})}\right) b_{p\da}-  \\ \nonumber
&& -e^{-i(\D\e_{k-p-q,k}-\om_p)t_1}e^{i(\D\e_{k-p-q,k}-\om_{p'})t_2} b^*_{p\da} \left(\frac{-e^{i(\D\e_{k-p+p'-q,k-p''}-\om_{p''})t_2}d\pd_{k-p'',k-p+p'-q,\da}b_{p''\da}}
{\gamma+i(\D\e_{k-p+p'-q,k-p''}-\om_{p''})}- \right. \\ \nonumber
&&\left. \left. - \frac{e^{i(\D\e_{k-p+p''-q,k-p'}-\om_{p''})t_2}d\pd_{k-p',k-p+p''-q,\da} b_{p''\da}}
{\gamma+i(\D\e_{k-p+p''-q,k-p'}-\om_{p''})}\right)b_{p'\da} \right] \\ \nonumber
\eea

\begin{figure}\centering
\includegraphics[scale=0.7]{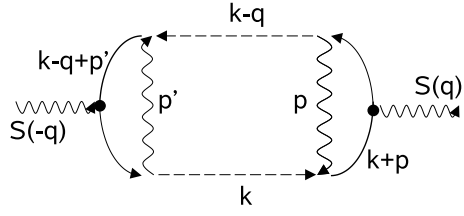}
\caption{Diagrammatic representation of the $1\times 3$ excitation process with correlated photons. The diagram shows the emitted spin-wave (wiggly lines to the left \& right of the diagram), pump light correlations (wiggly lines inside) and particle-hole excitations (solid \& dashed lines, respectively).}
\label{diagram-one-fermion-loop}
\end{figure}

The contraction $\la S_z^{(1)}(\bar{q},t)S_z^{(3)}(q,t)\ra$ leads to contractions such as

\be \la b^*_{\bar{p}} b^*_{p'} b_{p''} b_p \ra \la P_{\bar{k}-\bar{p}-\bar{q},\bar{k}}\pd_{k-p+p'-p'',k-q}\ra \ee

which result in the constraint $\bar{k}=k-q$  which restrict the phase space of $k,\bar{k}$ summation considerably, compared to the $2\times 2$ contribution. Diagrammatically, these fourth order contributions can be described as one fermion loop diagrams, see Fig. \ref{diagram-one-fermion-loop}.

\end{document}